\documentclass[english,twocolumn,aps,prb]{revtex4}
\usepackage[T1]{fontenc}
\usepackage[latin9]{inputenc}
\usepackage{units}
\usepackage{amstext}
\usepackage{graphicx}



\usepackage{babel}

\begin{document}

\title{Temperature and Voltage Driven Tunable Metal-Insulator Transition in Individual $W_{x}V_{1-x}O_{2}$ nanowires}

\author{Tai-Lung Wu$^{\text{1}}$}

\author{Luisa Whittaker$^{\text{2}}$}

\author{Sarbajit Banerjee$^{\text{2}}$}
\email{sb244@buffalo.edu}

\author{G.~Sambandamurthy$^{\text{1}}$}

\email{sg82@buffalo.edu}

\affiliation{$^{\text{1}}$Department of Physics and $^{\text{2}}$Department
of Chemistry, University at Buffalo, State University of New York, Buffalo, NY 14260, USA }
\begin{abstract}

Results from transport measurements in individual $W_{x}V_{1-x}O_{2}$
nanowires with varying extents of $W$ doping are presented. An abrupt thermally driven metal-insulator transition (MIT) is observed in these wires and the transition temperature decreases with increasing $W$ content at a pronounced rate of - (48-56) K/$at.\%W$, suggesting a significant alteration of the phase diagram from the bulk. These nanowires can also be driven through a voltage-driven MIT and the temperature dependence of the insulator to metal and metal to insulator switchings are studied. While driving from an insulator to metal, the threshold voltage at which the MIT occurs follows an exponential temperature dependence ($V_{TH\uparrow}\propto\exp(\nicefrac{-T}{T_{0}})) $whereas driving from a metal to insulator, the threshold voltage follows $V_{TH\downarrow}\propto\sqrt{T_{c}-T}$ and the implications of these results are discussed.

\end{abstract}
\maketitle

Strongly correlated electron materials with closely linked structural, electronic and spin degrees of freedom exhibit novel transport and magnetic properties such as half-metallicity, high $T_c$ superconductivity, colossal magnetoresistance (CMR), ferroelectricity, and metal-insulator transition (MIT). The understanding of and the control over the underlying microscopic mechanisms responsible for these macroscopic phenomena have been long standing research problems.\cite{2005EDagotto,2006PALee,1998MImada} When one or more dimensions of a strongly correlated material is reduced, several unexpected physical phenomena emerge. Recent observations of nanoscale phase separation in CMR oxides and in vanadium oxide films are examples of such correlated behavior at the nanoscale.\cite{2010KLai,2007MMQazilbash} In this paper, we study the doping and confinement effects on the transport properties of a strongly correlated system, $W_{x}V_{1-x}O_{2}$, at the nanoscale.

Vanadium oxide ($VO_{2}$), a well-known model system to study MIT and electron correlation over several decades, continues to provide many new and exciting results such as phase nucleation in individual nanobeams and thin films, multiple avalanches across the transition in confined films, and evidence for ultrafast resistance switching at sub-picosecond timescales.\cite{2007MMQazilbash,2008ASharoni,2001ACavalleri} When $VO_{2}$ is heated to $T\sim$340 K, it undergoes an orders-of-magnitude drop in electrical resistance from a high-resistance insulating phase (I) to a low-resistance metallic phase (M).\cite{1959FJMorin} This hysteretic MIT is also accompanied by a Peierl's type phase transition from a low-$T$ monoclinical phase to a high-$T$ tetragonal phase accompanied by an increase of symmetry and disruption of $V$-$V$ dimers along the monoclinic $c$-axis.\cite{1971JBGoodenough,1998MImada} Besides this thermally induced transition, MIT in $VO_{2}$ systems can also be driven electrically, optically or by external strain.\cite{2010HTKim,2008MRini,2009JCao} The origin of this pronounced phase transition has been attributed variously to Peierls instability driven by strong electron-phonon coupling as well as to Coulombic repulsion and electron localization due to electron-electron interactions that are described by Mott-Hubbard picture.\cite{1968NFMott,1971JBGoodenough,1994RMWentzcovitch,1994TMRice}

Recently, considerable research has been done on understanding the influence of finite size in altering the phase transition and hysteresis of $VO_{2}$.\cite{2008ASharoni,2009SZhang,2009LWhittaker} A recent scanning near-field IR microscopy study on thin $VO_{2}$ films clearly shows nucleation of metal\textendash{}insulator domains around the transition $T$ \cite{2007MMQazilbash} and  similar domains are also reported in  $VO_{2}$ nanobeams.\cite{2009JWei,2009JCao} Due to the generally abrupt nature of the MIT, $VO_{2}$ has been suggested as a potential candidate for Mott field-effect transistors, thermochromic window coatings and ultrafast optical shutter devices, infrared waveguides and modulators especially in reduced dimensions.\cite{1997CZhou,2010SHormoz,2004TDManning,2008TBenMessaoud,2010RMBriggs} Hence, studies of individual nanostructures are particularly important to understand single-domain phenomena when the dimensions of the system approach the intrinsic domain size.\cite{2008ASharoni,2009JCao,2009SZhang,2009JWei}

Our goals in this study are three-fold: (1) to understand the effect of $W$ doping on the transport properties of individual, single-crystalline nanowires of $VO_{2}$, (2) to understand the effect of finite size in influencing the MIT in doped nanowires and (3) to discern the similarities and/or differences in the microscopic transport mechanisms from metal to insulator (M$\rightarrow$I) and insulator to metal (I$\rightarrow$M) transitions in these nanostructures. Substitutional doping of both the $V$ and $O$ sublattices, for example by $Cr$, $Mo$ or $W$, has been extensively used in thin films as routes for controlling the phase transition temperature.\cite{1974DBMcWhan,1975AZylbersztejn,1985CTang,2009KLHolman,2007QGu} In particular, doping the $V$ sublattice with $W$ is very attractive since it yields a pronounced reduction in the transition $T$ and it structurally stabilizes the tetragonal phase.\cite{1985CTang,2007QGu,2007CKim,2010KShibuya,2010BGChae,2010LWhittaker}

\begin{figure}
\includegraphics[width=1\columnwidth]{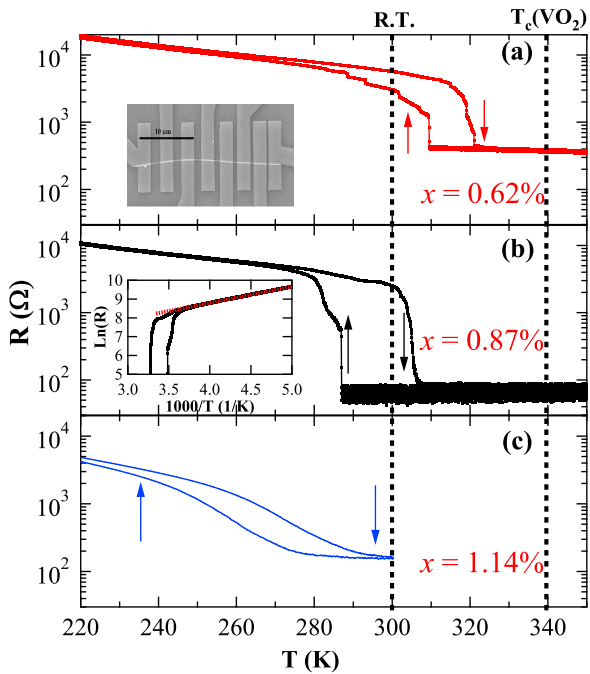} \caption{Thermally induced MIT in three representative nanowires with varying extents of $W$ doping of (a) 0.62\%, (b) 0.87\%, and (c) 1.14\% . Arrows indicate the sweep direction and the vertical dashed lines represent
the $T_{c}$ of undoped, bulk $VO_{2}$ and room $T$. Inset (a): SEM image of a typical $W_{x}V_{1-x}O_{2}$ nanowire device. All the results presented are for 5 $\mu$m long devices. Inset (b): $ln(R)\,-\,$1000/T plot follows a linear behavior in the low $T$ region showing thermally activated behavior.}
\label{fig1}
\end{figure}

The $W_{x}V_{1-x}O_{2}$ nanowires in this study are synthesized by the hydrothermal reduction of bulk $V_{2}O_{5}$ in the presence of small-molecule reducing agents with tungstic acid as the dopant precursor. Detailed description about the synthesis and structural characterization of our nanowires can be found elsewhere.\cite{2010LWhittaker} The $W$ dopant composition (atomic percentage of $W$, $x$) of our $W_{x}V_{1-x}O_{2}$ nanowires is found to be in the range of 0.25$-$1.14\% by ICP-MS analysis. The nanowires studied here range in length from 0.4-20 $\mu$m and have rectangular cross-sections with widths on the order of 400-600 nm and heights that show considerably greater confinement with dimensions between 15-100 nm. Nanowires are first dispersed and diluted in isopropanol by ultrasonication and then spun onto a Si/SiO$_{\text{2}}$ substrate. On top of each individual nanowire, Cr/Au electrodes are patterned and deposited by standard lithography and using an electron beam evaporator respectively. The inset of Fig. \ref{fig1}(a) shows one of our typical $W_{x}V_{1-x}O_{2}$ nanowire device.  In this work,  we only present results from devices with a length of 5 $\mu$m. The resistance measurements are performed using standard low frequency lock-in techniques and/or d.c. techniques with excitation currents of $\leq$ 50 nA. $T$ is controlled with a sufficiently slow scanning rate ($\leq$ 3 K/min) in each of our measurements.

Figure. \ref{fig1} shows the $T$ dependence of resistance ($R$) for three of our nanowires with different $W$-compositions ($x$) when they are driven through the MIT. We define the transition temperature ($T_{c}$) to be the mid point of  the heating cycle curve in our samples.
For an undoped $VO_{2}$ sample, $T_{c}$ $\sim$ 341 K whereas increased extent of $W$ doping progressively depresses $T_{c}$ until at $x$ = 1.14, the transition regime is entirely depressed below 300 K, as seen in Fig. 1(c). This is important for several applications such as energy saving window coatings that require room temperature operation.\cite{2004TDManning} In addition to this controlled reduction of $T_{c}$ with doping, another curious aspect of the transitions is seen in Fig. \ref{fig1}.  Multiple discontinuous jumps in $R$ in the lower composition nanowires (0.62\% and 0.87\%) imply that metal\textendash{}insulator domains co-exist around the transition region as observed recently in $VO_{2}$ nanostructures.\cite{2008ASharoni,2007MMQazilbash} The observation of these avalanche features may suggest the intermediacy of the antiferromagnetic $M_{2}$ phase, which is known to be stabilized by inhomogeneous tensile strain and doping, both features likely being manifested in our nanowires.\cite{2009SZhang} Additionally, coupling of the substrate to the high-surface-area nanowires imposes inhomogeneous stresses along the nanowire growth direction, resulting in formation of periodic insulating and metallic domains that propagate and change in size with increasing $T$  under the counteracting influences of the domain wall energy and relief of elastic strain.\cite{2009JWei} Interestingly, these abrupt steps disappear when $x$ increases to 1.14\%. Control over these  avalanches by $W$ doping may also be a potentially useful feature for certain applications.

\begin{figure}[b]
\includegraphics[width=0.75\columnwidth]{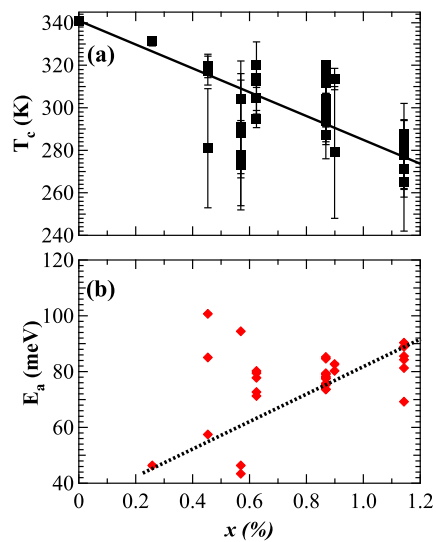} \caption{(a) The plot of transition temperature ($T_{c}$) as function of $W$ composition ($x$) for all of our nanowires. $T_{c}$ values are effectively reduced by $W$ doping at  $\sim$ -(48-56) K/$at.\%W$. (b) Thermal activation energy ($E_{a}$) in insulating phase as a function of $W$ composition. The lines are guides to eye.}
\label{fig2}
\end{figure}

\begin{figure*}[t]
\includegraphics[scale=0.75]{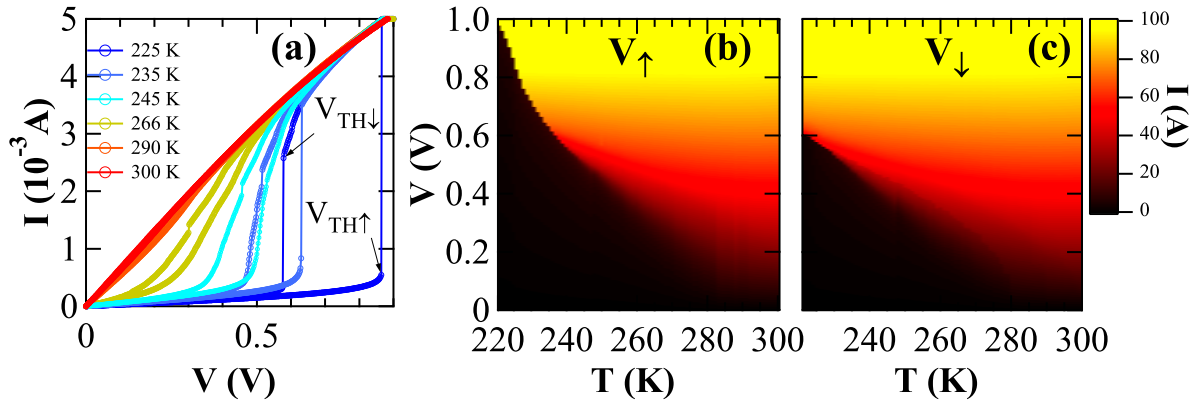} \caption{(a) Current-Voltage characteristics of a single $W_{x}V_{1-x}O_{2}$ ($x=1.14\%$) nanowire, taken at different $T$s, showing the V-driven MIT. Color maps of the current through the sample are plotted in the $V$ - $T$ plane when driving the sample through the (b) I$\rightarrow$M  and (c) M$\rightarrow$I  transitions respectively.}
\label{fig3}
\end{figure*}

Figure \ref{fig2}(a) shows the collection of  $T_{c}$ values as a function of $W$ composition ($x$) for several of our nanowires with $x$ values ranging from 0.26-1.14\%. It is clear that $W$ doping in single nanowires is an effective way to reduce $T_{c}$ and we can now engineer nanowires with $T_{c}$ values between 260 and 340 K, by controlling the amount of $W$.  $T_{c}$ decreases with increasing $x$ and shows an approximately linear relation with a slope in the range -(48-56) K/$at.\%W$. This value is dramatically larger than observed previously, such as a -18.4 K/$at.\%W$ in nanobeams (Ref. \onlinecite{2007QGu}) and the -10 to -27.8 K/$at.\%W$ in thin films (Ref. \onlinecite{2007CKim,2010BGChae,2007OBerezina}). The much more pronounced depression of $T_{c}$ upon $W$ doping observed for the nanowires suggests a significant alteration of the phase diagram from the bulk. Since nanowires can support single domains along their confined dimensions (the width and height), it is likely that substitutional $W$-doping and concomitant nucleation of the tetragonal phase has a much stronger influence in the nanowires wherein domains need to propagate only in one dimension as compared to bulk or thin film samples. The MIT is hysteretic, as  is typical of a first-order transition; the hysteresis in our nanowires is $\sim$10 - 40 K and the hysteresis gap does not show any correlation with $x$ in the range of 0.26-1.14\%.

Resistance of our nanowires in the low $T$ insulating phase follows an Arrhenious behavior, $R(T)=R_{0}\cdot\exp[E_{a}/kT]$.  Fitting to the Arrhenious form (see Fig. \ref{fig1}(b) inset), we extract the thermal activation energy ($E_{a}$) and our nanowires have values ranging from 40 - 100 meV. Figure \ref{fig2}(b) summarizes the $E_{a}$ values for all our nanowires as a function of $x$. Interestingly, $E_{a}$ values, though somewhat scattered, has a general trend of increasing with $x$. The $E_{a}$ values in this study are lower than the 300 meV observed in the suspended $VO_{2}$ nanobeams (Ref. \onlinecite{2009JWei}) that corresponds to the intrinsic optical bangap of $VO_{2}$. However, our data is comparable to the 90 meV in $VO_{2}$ nanobeam and the 82 meV in the $W_{0.003}V_{0.997}O_{2}$ film studies (Ref. \onlinecite{2009SZhang,2010BGChae}) and may be suggestive of more complex domain dynamics in doped nanobeams because of the induction of localized distortions by the substitutional dopant atoms. At low doping concentrations, the measured activation energies are consistent with the reported bandgap of the  intermediate $M_{2}$ phase.\cite{2009SZhang}

Having established control over the transport properties of single nanowires of $W$-doped $VO_{2}$, we now turn our attention to our main result. As mentioned earlier, MIT in $W_{x}V_{1-x}O_{2}$ single-crystalline nanowires can also driven by applying a voltage ($V$) across the source-drain contacts of the device. $V$ (or electric field)-driven MIT is extremely useful in several applications where thermal-driven transitions are not possible.\cite{1997CZhou,2010SHormoz}

At $T<T_{C}$, for example in the trace taken at 225 K in Fig. \ref{fig3}(a), the current increases slowly when increasing $V$ with a slope corresponding to $R$ in the insulating phase. When $V$ reaches a threshold voltage ($V_{TH\uparrow}$), the current ($I$) through the sample shows a discontinuous jump, indicating a I$\rightarrow$M transition. Beyond $V_{TH\uparrow}$, all $I$-$V$ curves show ohmic behavior. Upon reducing $V$ in the low-$R$ metallic phase, $I$ decreases until a second, albeit lower $V_{TH\downarrow}$ when a second discontinuous jump marking the return to the insulating phase is noticed.\cite{2009GGopalakrishnan,2008JMBaik,2008BJKim,2008CKo2,2004HTKim} It is interesting to note that when approaching $V_{TH\downarrow}$, a series of small current jumps are observed though the $R$-$T$ curve of this particular sample in Fig. \ref{fig1}(c) is smooth suggesting the intermediacy of the $M_{2}$ phase or inhomogeneous compression induced transient stabilization of multiple domains.

It is clear from Fig. \ref{fig3}(a) that $V_{TH\uparrow}\neq V_{TH\downarrow}$ and hysteresis is seen suggesting possibly that the underlying mechanisms behind I$\rightarrow$M  and M$\rightarrow$I transition could be different. These discontinuous $I$-$V$ curves and hysteresis persists till we approach closer to $T_{c}$ when the transition becomes smoother.
Before we discuss the implications of this observation, it is important to construct a map of the M$\rightarrow$I and I$\rightarrow$M transition in the $V_{TH}$ - $T$ phase space for our nanowires. In Fig. \ref{fig3}(b) and (c) the behavior of our samples is summarized in the form of a 2D color map of the $I$ values in the $V_{\uparrow}$ - $T$ and $V_{\downarrow}$ - $T$ planes respectively. Sharp transition points can be seen in the map as a sudden change in color for $T$ below $\sim$ 250 K in both panels. The $T$ dependence of these transition points is discussed below.

\begin{figure}[h]
\includegraphics[width=.75\columnwidth]{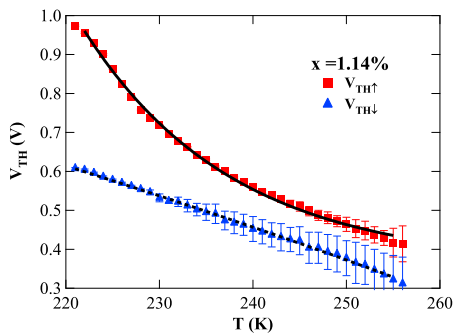} \caption{$V_{TH}-T$ plot for both driving directions. The solid line is the fitting curve for $V_{TH\uparrow}$ with the equation ($V_{TH\uparrow}\propto\exp(\nicefrac{-T}{T_{0}})$ ). The dashed line is the fitting curve for $V_{TH\downarrow}$ with the equation ($V_{TH\downarrow}\propto\sqrt{T_{c}-T}$).}
\label{fig4}
\end{figure}

In Fig. \ref{fig4}, we plot both the threshold voltages ($V_{TH\uparrow}$ and $V_{TH\downarrow}$ ) as function of $T$. When driving from I$\rightarrow$M, the $T$ dependence of $V_{TH\uparrow}$ has the exponential form ($V_{TH\uparrow}\propto\exp(\nicefrac{-T}{T_{0}}))$, as seen by the solid line in Fig. \ref{fig4}. The insulator (M1) in undoped $VO_{2}$ is thought of as a Mott insulator with charge ordering that is likely caused by electron-phonon interactons.\cite{2007MMQazilbash} In a charge ordered system, an exponential $T$ dependence of $V_{TH\uparrow}$ has often been found (Ref. \onlinecite{2000YTaguchi,1986KMaki}) and our result indicates that the $V$-driven I$\rightarrow$M transition in $W_{x}V_{1-x}O_{2}$ nanowires follows the same $T$ dependence.
In the opposite driving direction (M$\rightarrow$I), the initial high current through the nanowire generates an increased heating. Assuming that the heat dissipation is entirely through Joule heating due to scattering of charge carriers in the metallic phase, then the $T$ dependence of $V_{TH\downarrow}$ for the M$\rightarrow$I transition can be expressed as $V_{TH\downarrow}\propto\sqrt{T_{c}-T}$ (Ref. \onlinecite{1975FAChudnovskii}), as shown by the dashed line in Fig. \ref{fig4}. In other words, the availability of additional carrier concentration during the metal-insulator phase transition results in the observed asymmetry between the two transitions.\cite{2008HTKim}

In summary, thermal and voltage driven MITs are observed in individual single-crystalline nanowires of $W$-doped $VO_{2}$. By controlling the $W$ composition, a tunable transition temperature ($T_{c}$) and thermal activation energy ($E_{a}$) can be achieved in our nanowires. We also show that the $T$ dependence of the threshold voltage ($V_{TH}$) in I-V measurements can be explained by charge ordering for the I$\rightarrow$M transition and Joule heating model for the M$\rightarrow$I transition. We thank Dr. Zeng's group for allowing us access to their PPMS system. This work is supported by the UB-SUNY's IRDF grant and NSF-DMR grants (0847169 and 0847324)

\bibliographystyle{apsrev4-1}

\begin{thebibliography}{42}
\expandafter\ifx\csname natexlab\endcsname\relax\def\natexlab#1{#1}\fi
\expandafter\ifx\csname bibnamefont\endcsname\relax
  \def\bibnamefont#1{#1}\fi
\expandafter\ifx\csname bibfnamefont\endcsname\relax
  \def\bibfnamefont#1{#1}\fi
\expandafter\ifx\csname citenamefont\endcsname\relax
  \def\citenamefont#1{#1}\fi
\expandafter\ifx\csname url\endcsname\relax
  \def\url#1{\texttt{#1}}\fi
\expandafter\ifx\csname urlprefix\endcsname\relax\def\urlprefix{URL }\fi
\providecommand{\bibinfo}[2]{#2}
\providecommand{\eprint}[2][]{\url{#2}}

\bibitem[{\citenamefont{Dagotto}(2005)}]{2005EDagotto}
\bibinfo{author}{\bibfnamefont{E.}~\bibnamefont{Dagotto}},
  \bibinfo{journal}{Science} \textbf{\bibinfo{volume}{309}},
  \bibinfo{pages}{257} (\bibinfo{year}{2005}).

\bibitem[{\citenamefont{Lee et~al.}(2006)\citenamefont{Lee, Nagaosa, and
  Wen}}]{2006PALee}
\bibinfo{author}{\bibfnamefont{P.~A.} \bibnamefont{Lee}},
  \bibinfo{author}{\bibfnamefont{N.}~\bibnamefont{Nagaosa}}, \bibnamefont{and}
  \bibinfo{author}{\bibfnamefont{X.-G.} \bibnamefont{Wen}},
  \bibinfo{journal}{Rev. Mod. Phys.} \textbf{\bibinfo{volume}{78}},
  \bibinfo{pages}{17} (\bibinfo{year}{2006}).

\bibitem[{\citenamefont{Imada et~al.}(1998)\citenamefont{Imada, Fujimori, and
  Tokura}}]{1998MImada}
\bibinfo{author}{\bibfnamefont{M.}~\bibnamefont{Imada}},
  \bibinfo{author}{\bibfnamefont{A.}~\bibnamefont{Fujimori}}, \bibnamefont{and}
  \bibinfo{author}{\bibfnamefont{Y.}~\bibnamefont{Tokura}},
  \bibinfo{journal}{Rev. Mod. Phys.} \textbf{\bibinfo{volume}{70}},
  \bibinfo{pages}{1039} (\bibinfo{year}{1998}).

\bibitem[{\citenamefont{Lai et~al.}(2010)}]{2010KLai}
\bibinfo{author}{\bibfnamefont{K.}~\bibnamefont{Lai}} \bibnamefont{\textit{et~al.}},
  \bibinfo{journal}{Science} \textbf{\bibinfo{volume}{329}},
  \bibinfo{pages}{190} (\bibinfo{year}{2010}).

\bibitem[{\citenamefont{Qazilbash et~al.}(2007)}]{2007MMQazilbash}
\bibinfo{author}{\bibfnamefont{M.~M.} \bibnamefont{Qazilbash}}
  \bibnamefont{\textit{et~al.}}, \bibinfo{journal}{Science}
  \textbf{\bibinfo{volume}{318}}, \bibinfo{pages}{1750} (\bibinfo{year}{2007}).

\bibitem[{\citenamefont{Sharoni et~al.}(2008)\citenamefont{Sharoni, Ramirez,
  and Schuller}}]{2008ASharoni}
\bibinfo{author}{\bibfnamefont{A.}~\bibnamefont{Sharoni}},
  \bibinfo{author}{\bibfnamefont{J.~G.} \bibnamefont{Ramirez}},
  \bibnamefont{and} \bibinfo{author}{\bibfnamefont{I.~K.}
  \bibnamefont{Schuller}}, \bibinfo{journal}{Phys. Rev. Lett.}
  \textbf{\bibinfo{volume}{101}}, \bibinfo{eid}{026404} (\bibinfo{year}{2008}).

\bibitem[{\citenamefont{Cavalleri et~al.}(2001)}]{2001ACavalleri}
\bibinfo{author}{\bibfnamefont{A.}~\bibnamefont{Cavalleri}}
  \bibnamefont{\textit{et~al.}}, \bibinfo{journal}{Phys. Rev. Lett.}
  \textbf{\bibinfo{volume}{87}}, \bibinfo{pages}{237401}
  (\bibinfo{year}{2001}).

\bibitem[{\citenamefont{Morin}(1959)}]{1959FJMorin}
\bibinfo{author}{\bibfnamefont{F.~J.} \bibnamefont{Morin}},
  \bibinfo{journal}{Phys. Rev. Lett.} \textbf{\bibinfo{volume}{3}},
  \bibinfo{pages}{34} (\bibinfo{year}{1959}).

\bibitem[{\citenamefont{Goodenough}(1971)}]{1971JBGoodenough}
\bibinfo{author}{\bibfnamefont{J.~B.} \bibnamefont{Goodenough}},
  \bibinfo{journal}{J. Solid State Chem.} \textbf{\bibinfo{volume}{3}},
  \bibinfo{pages}{490} (\bibinfo{year}{1971}).

\bibitem[{\citenamefont{Kim et~al.}(2010)}]{2010HTKim}
\bibinfo{author}{\bibfnamefont{H.~T.} \bibnamefont{Kim}} \bibnamefont{\textit{et~al.}},
  \bibinfo{journal}{J. Appl. Phys.} \textbf{\bibinfo{volume}{107}},
  \bibinfo{pages}{023702} (\bibinfo{year}{2010}).

\bibitem[{\citenamefont{Rini et~al.}(2008)}]{2008MRini}
\bibinfo{author}{\bibfnamefont{M.}~\bibnamefont{Rini}} \bibnamefont{\textit{et~al.}},
  \bibinfo{journal}{Appl. Phys. Lett.} \textbf{\bibinfo{volume}{92}},
  \bibinfo{eid}{181904} (\bibinfo{year}{2008}).

\bibitem[{\citenamefont{Cao et~al.}(2009)}]{2009JCao}
\bibinfo{author}{\bibfnamefont{J.}~\bibnamefont{Cao}} \bibnamefont{\textit{et~al.}},
  \bibinfo{journal}{Nature Nanotech.} \textbf{\bibinfo{volume}{4}},
  \bibinfo{pages}{732} (\bibinfo{year}{2009}).

\bibitem[{\citenamefont{Mott}(1968)}]{1968NFMott}
\bibinfo{author}{\bibfnamefont{N.~F.} \bibnamefont{Mott}},
  \bibinfo{journal}{Rev. Mod. Phys.} \textbf{\bibinfo{volume}{40}},
  \bibinfo{pages}{677} (\bibinfo{year}{1968}).

\bibitem[{\citenamefont{Wentzcovitch et~al.}(1994)\citenamefont{Wentzcovitch,
  Schulz, and Allen}}]{1994RMWentzcovitch}
\bibinfo{author}{\bibfnamefont{R.~M.} \bibnamefont{Wentzcovitch}},
  \bibinfo{author}{\bibfnamefont{W.~W.} \bibnamefont{Schulz}},
  \bibnamefont{and} \bibinfo{author}{\bibfnamefont{P.~B.} \bibnamefont{Allen}},
  \bibinfo{journal}{Phys. Rev. Lett.} \textbf{\bibinfo{volume}{72}},
  \bibinfo{pages}{3389} (\bibinfo{year}{1994}).

\bibitem[{\citenamefont{Rice et~al.}(1994)\citenamefont{Rice, Launois, and
  Pouget}}]{1994TMRice}
\bibinfo{author}{\bibfnamefont{T.~M.} \bibnamefont{Rice}},
  \bibinfo{author}{\bibfnamefont{H.}~\bibnamefont{Launois}}, \bibnamefont{and}
  \bibinfo{author}{\bibfnamefont{J.~P.} \bibnamefont{Pouget}},
  \bibinfo{journal}{Phys. Rev. Lett.} \textbf{\bibinfo{volume}{73}},
  \bibinfo{pages}{3042} (\bibinfo{year}{1994}).

\bibitem[{\citenamefont{Zhang et~al.}(2009)\citenamefont{Zhang, Chou, and
  Lauhon}}]{2009SZhang}
\bibinfo{author}{\bibfnamefont{S.}~\bibnamefont{Zhang}},
  \bibinfo{author}{\bibfnamefont{J.~Y.} \bibnamefont{Chou}}, \bibnamefont{and}
  \bibinfo{author}{\bibfnamefont{L.~J.} \bibnamefont{Lauhon}},
  \bibinfo{journal}{Nano Lett.} \textbf{\bibinfo{volume}{9}},
  \bibinfo{pages}{4527} (\bibinfo{year}{2009}).

\bibitem[{\citenamefont{Whittaker et~al.}(2009)}]{2009LWhittaker}
\bibinfo{author}{\bibfnamefont{L.}~\bibnamefont{Whittaker}}
  \bibnamefont{\textit{et~al.}}, \bibinfo{journal}{J. Am. Chem. Soc.}
  \textbf{\bibinfo{volume}{131}}, \bibinfo{pages}{8884} (\bibinfo{year}{2009}).

\bibitem[{\citenamefont{Wei et~al.}(2009)}]{2009JWei}
\bibinfo{author}{\bibfnamefont{J.}~\bibnamefont{Wei}} \bibnamefont{\textit{et~al.}},
  \bibinfo{journal}{Nature Nanotech.} \textbf{\bibinfo{volume}{4}},
  \bibinfo{pages}{420} (\bibinfo{year}{2009}).

\bibitem[{\citenamefont{Zhou and Newns}(1997)}]{1997CZhou}
\bibinfo{author}{\bibfnamefont{C.}~\bibnamefont{Zhou}} \bibnamefont{and}
  \bibinfo{author}{\bibfnamefont{D.}~\bibnamefont{Newns}},
  \bibinfo{journal}{Appl. Phys. Lett.} \textbf{\bibinfo{volume}{70}},
  \bibinfo{pages}{598} (\bibinfo{year}{1997}).

\bibitem[{\citenamefont{Hormoz and Ramanathan}(2010)}]{2010SHormoz}
\bibinfo{author}{\bibfnamefont{S.}~\bibnamefont{Hormoz}} \bibnamefont{and}
  \bibinfo{author}{\bibfnamefont{S.}~\bibnamefont{Ramanathan}},
  \bibinfo{journal}{Solid-State Electron.} \textbf{\bibinfo{volume}{54}},
  \bibinfo{pages}{654} (\bibinfo{year}{2010}).

\bibitem[{\citenamefont{Manning et~al.}(2004)}]{2004TDManning}
\bibinfo{author}{\bibfnamefont{T.~D.} \bibnamefont{Manning}}
  \bibnamefont{\textit{et~al.}}, \bibinfo{journal}{Chem. Mater.}
  \textbf{\bibinfo{volume}{16}}, \bibinfo{pages}{744} (\bibinfo{year}{2004}).

\bibitem[{\citenamefont{Ben-Messaoud et~al.}(2008)}]{2008TBenMessaoud}
\bibinfo{author}{\bibfnamefont{T.}~\bibnamefont{Ben-Messaoud}}
  \bibnamefont{\textit{et~al.}}, \bibinfo{journal}{Opt. Commun.}
  \textbf{\bibinfo{volume}{281}}, \bibinfo{pages}{6024} (\bibinfo{year}{2008}).

\bibitem[{\citenamefont{Briggs et~al.}(2010)\citenamefont{Briggs, Pryce, and
  Atwater}}]{2010RMBriggs}
\bibinfo{author}{\bibfnamefont{R.~M.} \bibnamefont{Briggs}},
  \bibinfo{author}{\bibfnamefont{I.~M.} \bibnamefont{Pryce}}, \bibnamefont{and}
  \bibinfo{author}{\bibfnamefont{H.~A.} \bibnamefont{Atwater}},
  \bibinfo{journal}{Opt. Express} \textbf{\bibinfo{volume}{18}},
  \bibinfo{pages}{11192} (\bibinfo{year}{2010}).

\bibitem[{\citenamefont{McWhan et~al.}(1974)}]{1974DBMcWhan}
\bibinfo{author}{\bibfnamefont{D.~B.} \bibnamefont{McWhan}}
  \bibnamefont{\textit{et~al.}}, \bibinfo{journal}{Phys. Rev. B}
  \textbf{\bibinfo{volume}{10}}, \bibinfo{pages}{490} (\bibinfo{year}{1974}).

\bibitem[{\citenamefont{Zylbersztejn and Mott}(1975)}]{1975AZylbersztejn}
\bibinfo{author}{\bibfnamefont{A.}~\bibnamefont{Zylbersztejn}}
  \bibnamefont{and} \bibinfo{author}{\bibfnamefont{N.~F.} \bibnamefont{Mott}},
  \bibinfo{journal}{Phys. Rev. B} \textbf{\bibinfo{volume}{11}},
  \bibinfo{pages}{4383} (\bibinfo{year}{1975}).

\bibitem[{\citenamefont{Tang et~al.}(1985)}]{1985CTang}
\bibinfo{author}{\bibfnamefont{C.}~\bibnamefont{Tang}} \bibnamefont{\textit{et~al.}},
  \bibinfo{journal}{Phys. Rev. B} \textbf{\bibinfo{volume}{31}},
  \bibinfo{pages}{1000} (\bibinfo{year}{1985}).

\bibitem[{\citenamefont{Holman et~al.}(2009)}]{2009KLHolman}
\bibinfo{author}{\bibfnamefont{K.~L.} \bibnamefont{Holman}}
  \bibnamefont{\textit{et~al.}}, \bibinfo{journal}{Phys. Rev. B}
  \textbf{\bibinfo{volume}{79}}, \bibinfo{pages}{245114}
  (\bibinfo{year}{2009}).

\bibitem[{\citenamefont{Gu et~al.}(2007)}]{2007QGu}
\bibinfo{author}{\bibfnamefont{Q.}~\bibnamefont{Gu}} \bibnamefont{\textit{et~al.}},
  \bibinfo{journal}{Nano Lett.} \textbf{\bibinfo{volume}{7}},
  \bibinfo{pages}{363} (\bibinfo{year}{2007}).

\bibitem[{\citenamefont{Kim et~al.}(2007)\citenamefont{Kim, Shin, and
  Ozaki}}]{2007CKim}
\bibinfo{author}{\bibfnamefont{C.}~\bibnamefont{Kim}},
  \bibinfo{author}{\bibfnamefont{J.~S.} \bibnamefont{Shin}}, \bibnamefont{and}
  \bibinfo{author}{\bibfnamefont{H.}~\bibnamefont{Ozaki}}, \bibinfo{journal}{J.
  Phys.: Condens. Matter} \textbf{\bibinfo{volume}{19}},
  \bibinfo{pages}{096007} (\bibinfo{year}{2007}).

\bibitem[{\citenamefont{Shibuya et~al.}(2010)\citenamefont{Shibuya, Kawasaki,
  and Tokura}}]{2010KShibuya}
\bibinfo{author}{\bibfnamefont{K.}~\bibnamefont{Shibuya}},
  \bibinfo{author}{\bibfnamefont{M.}~\bibnamefont{Kawasaki}}, \bibnamefont{and}
  \bibinfo{author}{\bibfnamefont{Y.}~\bibnamefont{Tokura}},
  \bibinfo{journal}{Appl. Phys. Lett.} \textbf{\bibinfo{volume}{96}},
  \bibinfo{pages}{022102} (\bibinfo{year}{2010}).

\bibitem[{\citenamefont{Chae and Kim}(2010)}]{2010BGChae}
\bibinfo{author}{\bibfnamefont{B.~G.} \bibnamefont{Chae}} \bibnamefont{and}
  \bibinfo{author}{\bibfnamefont{H.~T.} \bibnamefont{Kim}},
  \bibinfo{journal}{Physica B} \textbf{\bibinfo{volume}{405}},
  \bibinfo{pages}{663} (\bibinfo{year}{2010}).

\bibitem[{\citenamefont{Whittaker et~al.}(2010)}]{2010LWhittaker}
\bibinfo{author}{\bibfnamefont{L.}~\bibnamefont{Whittaker}}
  \bibnamefont{\textit{et~al.}} (\bibinfo{year}{2010}), \bibinfo{note}{submitted for
  publication}.

\bibitem[{\citenamefont{Berezina et~al.}(2007)}]{2007OBerezina}
\bibinfo{author}{\bibfnamefont{O.}~\bibnamefont{Berezina}}
  \bibnamefont{\textit{et~al.}}, \bibinfo{journal}{Inorg. Mater.}
  \textbf{\bibinfo{volume}{43}}, \bibinfo{pages}{505} (\bibinfo{year}{2007}).

\bibitem[{\citenamefont{Gopalakrishnan
  et~al.}(2009)\citenamefont{Gopalakrishnan, Ruzmetov, and
  Ramanathan}}]{2009GGopalakrishnan}
\bibinfo{author}{\bibfnamefont{G.}~\bibnamefont{Gopalakrishnan}},
  \bibinfo{author}{\bibfnamefont{D.}~\bibnamefont{Ruzmetov}}, \bibnamefont{and}
  \bibinfo{author}{\bibfnamefont{S.}~\bibnamefont{Ramanathan}},
  \bibinfo{journal}{J. Mater. Sci.} \textbf{\bibinfo{volume}{44}},
  \bibinfo{pages}{5345} (\bibinfo{year}{2009}).

\bibitem[{\citenamefont{Baik et~al.}(2008)}]{2008JMBaik}
\bibinfo{author}{\bibfnamefont{J.~M.} \bibnamefont{Baik}} \bibnamefont{\textit{et~al.}},
  \bibinfo{journal}{J. Phys. Chem. C} \textbf{\bibinfo{volume}{112}},
  \bibinfo{pages}{13328} (\bibinfo{year}{2008}).

\bibitem[{\citenamefont{Kim et~al.}(2008{\natexlab{a}})}]{2008BJKim}
\bibinfo{author}{\bibfnamefont{B.-J.} \bibnamefont{Kim}} \bibnamefont{\textit{et~al.}},
  \bibinfo{journal}{Phys. Rev. B} \textbf{\bibinfo{volume}{77}},
  \bibinfo{pages}{235401} (\bibinfo{year}{2008}{\natexlab{a}}).

\bibitem[{\citenamefont{Ko and Ramanathan}(2008)}]{2008CKo2}
\bibinfo{author}{\bibfnamefont{C.}~\bibnamefont{Ko}} \bibnamefont{and}
  \bibinfo{author}{\bibfnamefont{S.}~\bibnamefont{Ramanathan}},
  \bibinfo{journal}{Appl. Phys. Lett.} \textbf{\bibinfo{volume}{93}},
  \bibinfo{pages}{252101} (\bibinfo{year}{2008}).

\bibitem[{\citenamefont{Kim et~al.}(2004)}]{2004HTKim}
\bibinfo{author}{\bibfnamefont{H.~T.} \bibnamefont{Kim}} \bibnamefont{\textit{et~al.}},
  \bibinfo{journal}{New J. Phys.} \textbf{\bibinfo{volume}{6}},
  \bibinfo{pages}{52} (\bibinfo{year}{2004}).

\bibitem[{\citenamefont{Taguchi et~al.}(2000)\citenamefont{Taguchi, Matsumoto,
  and Tokura}}]{2000YTaguchi}
\bibinfo{author}{\bibfnamefont{Y.}~\bibnamefont{Taguchi}},
  \bibinfo{author}{\bibfnamefont{T.}~\bibnamefont{Matsumoto}},
  \bibnamefont{and} \bibinfo{author}{\bibfnamefont{Y.}~\bibnamefont{Tokura}},
  \bibinfo{journal}{Phys. Rev. B} \textbf{\bibinfo{volume}{62}},
  \bibinfo{pages}{7015} (\bibinfo{year}{2000}).

\bibitem[{\citenamefont{Maki}(1986)}]{1986KMaki}
\bibinfo{author}{\bibfnamefont{K.}~\bibnamefont{Maki}}, \bibinfo{journal}{Phys.
  Rev. B} \textbf{\bibinfo{volume}{33}}, \bibinfo{pages}{2852}
  (\bibinfo{year}{1986}).

\bibitem[{\citenamefont{Chudnovskii}(1975)}]{1975FAChudnovskii}
\bibinfo{author}{\bibfnamefont{F.~A.} \bibnamefont{Chudnovskii}},
  \bibinfo{journal}{Sov. Phys. Tech. Phys.} \textbf{\bibinfo{volume}{20}},
  \bibinfo{pages}{999} (\bibinfo{year}{1975}).

\bibitem[{\citenamefont{Kim et~al.}(2008{\natexlab{b}})}]{2008HTKim}
\bibinfo{author}{\bibfnamefont{H.~T.} \bibnamefont{Kim}} \bibnamefont{\textit{et~al.}},
  \bibinfo{journal}{Physica B} \textbf{\bibinfo{volume}{403}},
  \bibinfo{pages}{1434} (\bibinfo{year}{2008}{\natexlab{b}}).

\end{thebibliography}

\end{document}